\documentclass[12pt,cite]{article}
\usepackage{epsfig}
\newcommand {\nut} {$\nu_{\tau}$}
\newcommand {\num} {$\nu_{\mu}$}

\begin{document}
\begin{flushright}
DFF 348/01/2000
\end{flushright}
\vspace{0.5cm}

\begin{center}

{\Large \bf Extreme Energy $\nu_{\tau}$ Propagation Through the Earth}

\vspace{1.8cm}

F. Becattini, S. Bottai \footnote{e-mail: becattini@fi.infn.it, bottai@fi.infn.it}

\vspace{1.0cm}

{\it University of Florence and INFN Sezione di Firenze}\\
{\it Largo E. Fermi 2, I-50125}\\
{\it Firenze  (Italy)}\\[0.5cm]

\vspace{2.5cm}
\begin{center}
{\bf Abstract}
\end{center}
\vspace{1cm}

\begin{minipage}{14cm}
\baselineskip=12pt
\parindent=0.5cm

{\small 
The propagation of extremely energetic \nut's through the Earth is studied 
by means of a detailed Monte Carlo simulation. All major mechanisms of \nut 
interactions and $\tau$ energy loss as well as all its relevant decay modes 
are properly taken into account. The detection possibility of \nut's emerging 
from Earth in forthcoming neutrino telescopes is addressed.}

\end{minipage}

\vspace{3cm}
{\it To be published in Astroparticle Physics}
\end{center}

\newpage

\section{Introduction}

Cosmic neutrinos with energy $ > 10^{12}\,$ eV are a new and exciting 
observational window for cosmic ray physics and astrophysics. Indeed, 
many interesting models predict the existence of cosmic neutrinos with 
an energy up to $10^{18}-10^{20}\,$ eV ({\em extreme energy}), 
where charged particles are expected to be cut-off by the so called GZK 
mechanism~\cite{GZK}. The aforementioned threshold of $10^{12}\,$ eV is 
necessary for a detection of diffuse flux, because 
of the large atmospheric neutrino background.\\
The detection of such extreme energy neutrinos is challenging as it
demands a very large sensitive area. The experiments of the upcoming 
generation plan to use underwater-underice Cerenkov detectors~\cite{gai95} and 
large field-of-view atmospheric fluorescence detectors~\cite{lin97}.
One of the best evidence of cosmic neutrinos would be the 
detection of upstream showers/particles emerging from the Earth. For this 
kind of events,
the atmospheric muon and primary charged cosmic ray background would be 
completely suppressed. On the other hand, this signature can hardly be
observed at extreme energy because the rise of weak cross sections entails 
the opacity of the Earth with respect to neutrino
propagation~\cite{nau98,gand}.\\
The problem of neutrino propagation through the Earth can be understood
on the basis of neutrino interactions with matter.
For neutrino energy well above few GeV, interactions
with conventional matter are described by neutrino-nucleon
deep inelastic scattering and neutrino-electron scattering~\cite{gand}. 
The former can be classified in two main categories: charged current (CC)
$\nu_{l}\,N\rightarrow\l\,+hadrons$, where the neutrino disappears
and a charged lepton $l$ is created, and neutral current (NC)
$\nu_{l}\,N\rightarrow\nu_{l}\,+hadrons$, where the neutrino survives 
the interaction. The neutrino-electron scattering
cross section is several order of magnitude less than neutrino-nucleon
cross section and does not affect significantly neutrino propagation in
matter. The only exception is the $W^-$ resonant scattering for the 
reaction $\bar{\nu}_{e}\,e\rightarrow W^-+anything$, which, however, hardly
affects neutrino attenuation length in matter for $\bar{\nu}_{e}$ energy
in the range $2\cdot\,10^{15}$~eV$\leq\,E_{\nu}\,\leq\,2\cdot\,10^{16}$~eV~\cite{gand}.\\
The neutrino-nucleon total cross section is approximately $\sigma_{tot}\approx
10^{-37}$~cm$^2$ at $E_{\nu}=10$~GeV and rises roughly linearly with
energy. As a consequence, while in the GeV energy range the Earth
is completely transparent to neutrinos, it is expected to become opaque
for sufficently high energy. As to muon neutrinos, the CC interaction lenght 
inside Earth equals Earth's diameter for energy around $40$~TeV~\cite{gand}. 
In order to accurately describe the change in energy spectrum of a muon 
neutrino flux traversing the Earth, both the effect of neutrino-nucleon CC 
and NC must be properly taken into account. The problem of neutrino propagation 
in the Earth can be solved accurately either by Monte-Carlo simulation 
or by approximate iterative solution of the relevant transport 
equation~\cite{nau98}.\\
It has been recently pointed out~\cite{far97, hal98,bot99,Iyer} that the 
behaviour of $\tau$-neutrinos, 
whose existence should be guaranteed in a neutrino-oscillation scenario, 
should be significantly different from $\nu_{\mu}$ and $\nu_e$.
Whilst muon and electron neutrinos are practically absorbed after
one CC interaction, the $\tau$ lepton created by the $\nu_{\tau}$ CC scattering may decay 
in flight before losing too much energy, thereby generating a new $\nu_{\tau}$ with
comparable energy. Hence, ultra high energy $\tau$-neutrinos should emerge 
from the Earth instead of being absorbed. 
For a correct evaluation of energy spectrum of $\nu_{\tau}$'s emerging
from the Earth, one has to properly take into account $\nu_{\tau}$ interactions
as well as $\tau$ energy loss and decay. An analytical approach similar
to that used in~\cite{nau98} for $\nu_{\mu}$ has been proposed
by S. Iyer et al.~\cite{Iyer} 
neglecting $\tau$ energy loss. This approach holds as long as energy
does not exceed $10^{16}$ eV because above this energy $\tau$ interaction
lenght becomes comparable with $\tau$ decay lenght (see next section).\\ 
In the present work, a detailed Monte Carlo calculation of $\nu_{\tau}-\tau$ 
system propagation through the Earth has been performed for energy up to 
$10^{20}$~eV including the $\tau$ energy loss contribution, with special 
emphasis on the initial $\nu_{\tau}$ spectrum deformation as a function of the zenith 
angle of the emerging particle.
\section{Neutrino cross sections and charged lepton interactions}

The Montecarlo simulation has been performed following neutrinos and
charged leptons along their path inside the Earth. The Earth model
considered is the preliminary Earth model of ref.~\cite{dzi81}.
During particle propagation, the occurrence of interactions relevant for energy loss, decay
or leptons production has been simulated by a unidimensional approach. The only
secondary particles which have been followed are the \nut's produced by $\tau$
decays and $\tau$'s arising from \nut~charged current interactions.\

\begin{figure}
\begin{center}
\epsfig{file=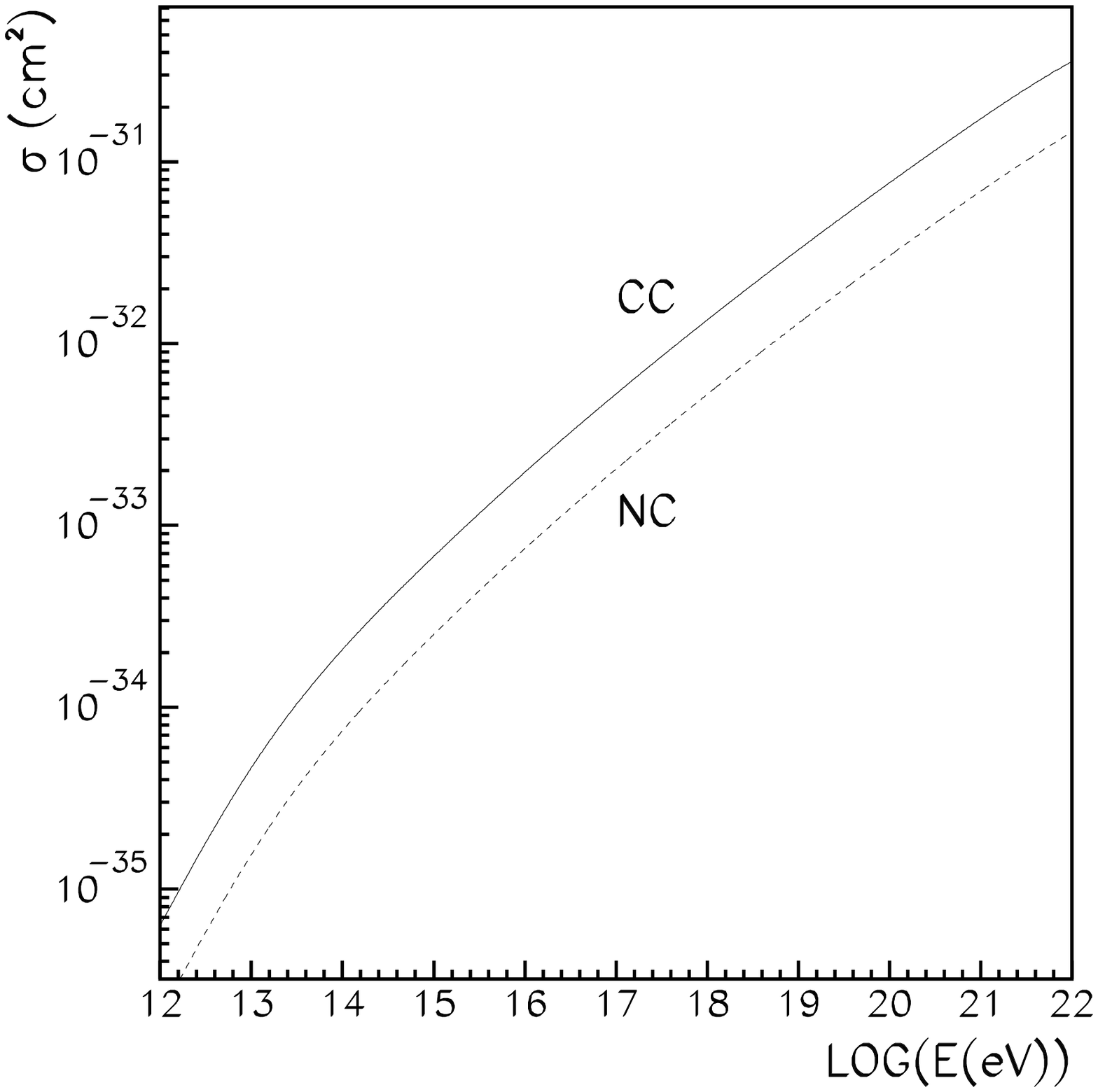,width=17cm}
\caption{Total \nut CC and NC cross sections on isoscalar nucleon target
as a function of energy.}
\end{center}
\label{fig:crossec}
\end{figure}

Deep inelastic neutrino-nucleon scattering is the dominant interaction of energetic 
neutrinos into conventional matter. Charged and neutral current differential cross 
sections used in this work have been calculated in the framework of QCD improved 
parton model. Charged current cross section for a neutrino-isoscalar nucleon 
interaction reads:

\begin{equation}
  \frac{d^2\sigma}{dxdy} = \frac{2 G_F^2 ME_\nu}{\pi} \left( 
  \frac{M_W^2}{Q^2 + M_W^2} \right)^{\!2} \left[xq(x,Q^2) + x 
  \bar{q}(x,Q^2)(1-y)^2 \right] , \label{eqn:sigsig}
\end{equation}
where $-Q^2$ is the invariant momentum transfer between the incident neutrino and 
outgoing muon, $\nu = E_\nu - E_l$ is the energy loss in the lab (target) frame, 
$x = Q^2/2M\nu$ and $y = \nu/E_\nu$, $M$ and $M_W$ are the nucleon and intermediate-boson 
masses, $G_F$ is the Fermi constant and the quark distribution functions are:

\begin{eqnarray}
 q(x,Q^2) & = & \frac{u_v(x,Q^2)+d_v(x,Q^2)}{2} + \frac{u_s(x,Q^2)+d_s(x,Q^2)}{2} + s_s(x,Q^2) + b_s(x,Q^2) \\
 \bar{q}(x,Q^2) & = & \frac{u_s(x,Q^2)+d_s(x,Q^2)}{2} + c_s(x,Q^2) + t_s(x,Q^2),\nonumber 
\end{eqnarray}
where the subscripts $v$ and $s$ label valence and sea contributions, and 
$u$, $d$, $c$, $s$, $t$, $b$ denote the distributions for various quark flavors.\\
Since a large contribution to cross sections comes from the very low $x$ region laying 
beyond present accelerator domain, an extrapolation of parton distribution at very low 
$x$ ($x\geq 10^{-8}$) is necessary. We have used the parton distribution set CTEQ3-DIS~\cite{lai}, 
available in the program library PDFLIB~\cite{plo97}, with NLO-DGLAP formalism used for 
$Q^2$ evolution and assuming for very low $x$ the same functional form measured at 
$x=\cal{O}$($10^{-5}$). Even if more sofisticated approaches for low $x$ extrapolation 
have been developed using dynamical QCD~\cite{glu98} the results for cross section 
calculations (see Fig.~\ref{fig:crossec}) do not differ more than $10\%$ from the approach 
taken here with CTEQ3-DIS plus "brute force" extrapolation~\cite{gand,glu98}.\\
The electromagnetic interactions of muons with matter, at the 
actually considered energies, are dominated by radiative processes 
rather than ionisation~\cite{lipa}. The cross sections for electromagnetic
radiative processes of $\tau$ are lower than muon's, still radiative 
interactions remain the dominant process for $\tau$ energy loss.
The cross sections used for radiative electromagnetic interactions of
$\tau$ leptons are based on QED calculation for
Bremsstrahlung~\cite{pet68}, for 
direct pair production~\cite{kok70} and for photonuclear
interactions~\cite{bez81} by replacing muon or electron mass with 
$\tau$ mass (only formulae with explicit lepton mass dependence in the
original paper have been considered). For all above processes 
we have implemented stochastic interactions for $\nu=(E_{f}-E_{i})/E_{i}\geq 10^{-3}$ 
and a continuous energy loss for $\nu=(E_{f}-E_{i})/E_{i}\le 10^{-3}$, where $E_{i}$ 
and $E_{f}$ are the $\tau$ energies before and after the interaction respectively.\\
It is worth mentioning that Bremsstrahlung cross section scales as the inverse 
square of lepton mass whereas the direct pair production approximately scales according 
to $m_e/m_l$~\cite{tan91}. As a consequence, the dominant process of $\tau$ lepton 
energy loss is direct pair production rather than photon radiation. Along with 
electromagnetic interactions, we also considered weak interactions of $\tau$ with 
nucleons although they are almost neglegible even at $E_{\tau}\sim10^{20}\,$ eV. 
The $\tau$ decay has been simulated by using the TAUOLA package~\cite{jad93}.

\section{Simulation results}

\begin{figure}
\begin{center}
\epsfig{figure=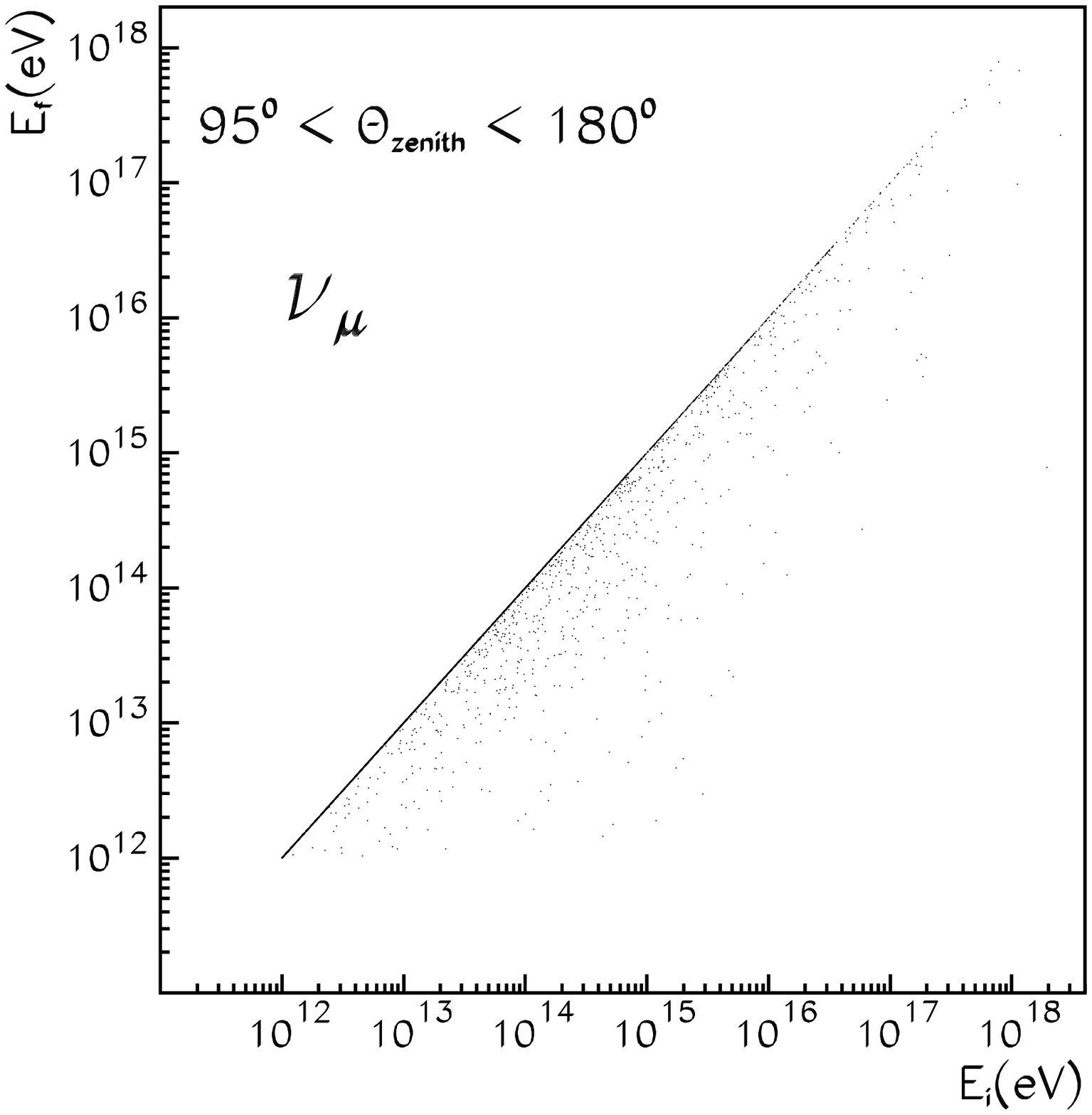,width=17cm}
\caption{Scatter plot of initial versus final energy of \num propagating
through the Earth with an incident zenith angle uniformly distributed in the 
range $95^0\leq\theta_{zenith}\leq 180^0$.}
\label{fig:scpl1}
\end{center}
\end{figure}

\begin{figure}
\begin{center}
\epsfig{figure=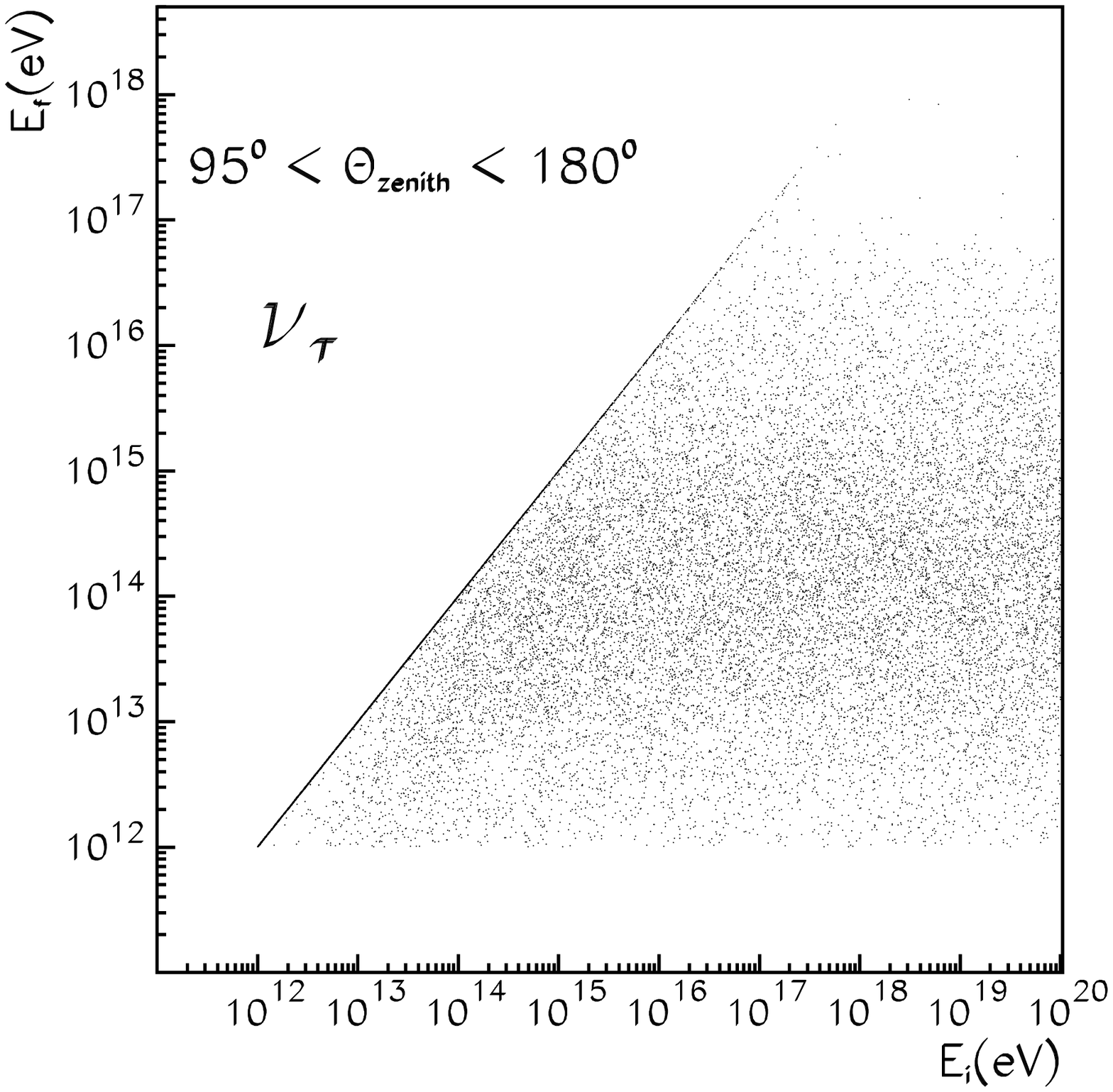,width=17cm}
\caption{Scatter plot of initial versus final energy of \nut propagating
through the Earth with an incident zenith angle uniformly distributed in the 
range $95^0\leq\theta_{zenith}\leq 180^0$.}
\label{fig:scpl2}
\end{center}
\end{figure}

\begin{figure}
\begin{center}
\epsfig{figure=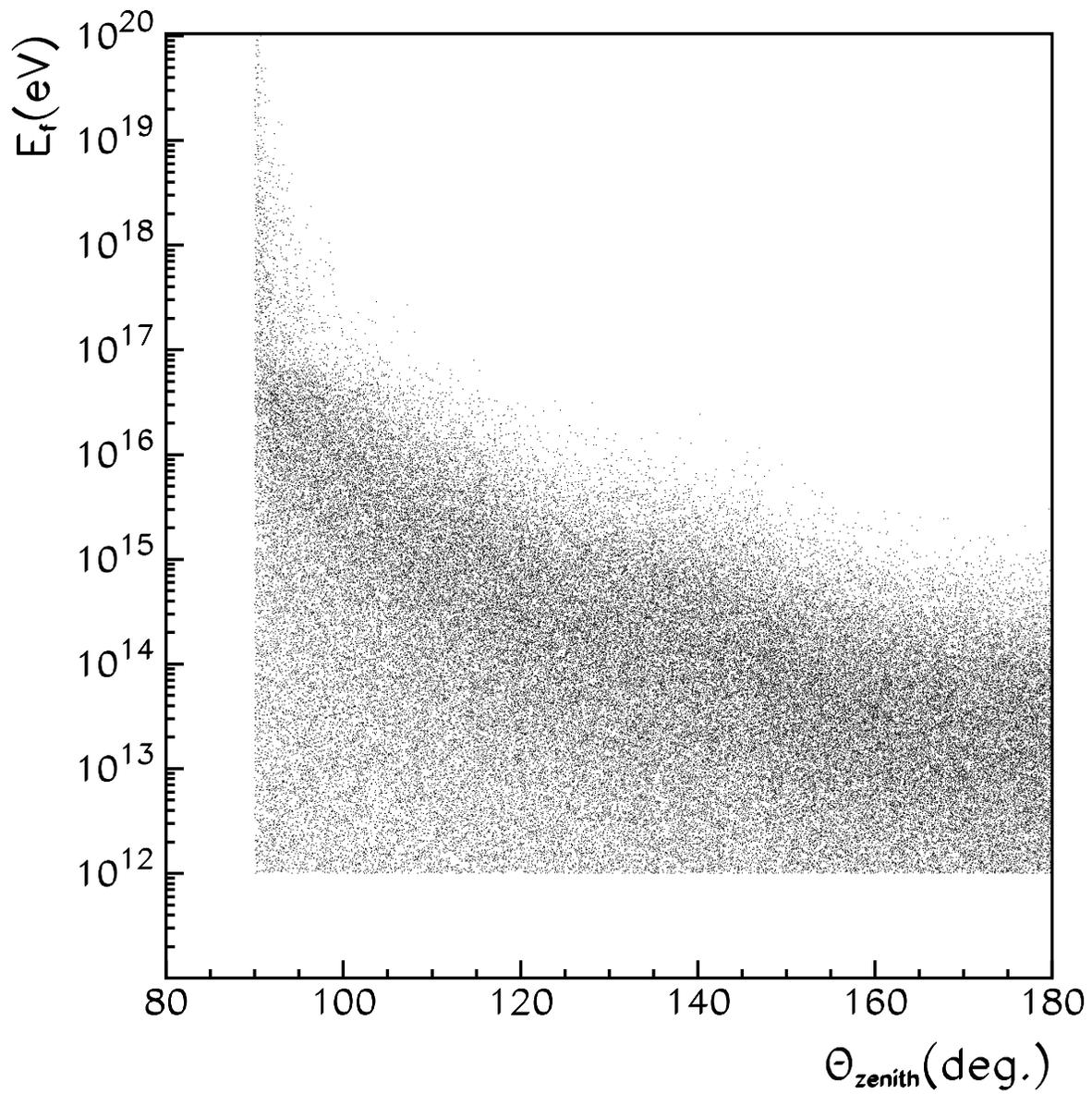,width=17cm}
\caption{Scatter plot of final energy as a function of zenith angle for
an initial energy spectrum $\propto E^{-1}$.}
\label{fig:scpl3}
\end{center}
\end{figure}

\begin{figure}
\begin{center}
\epsfig{figure=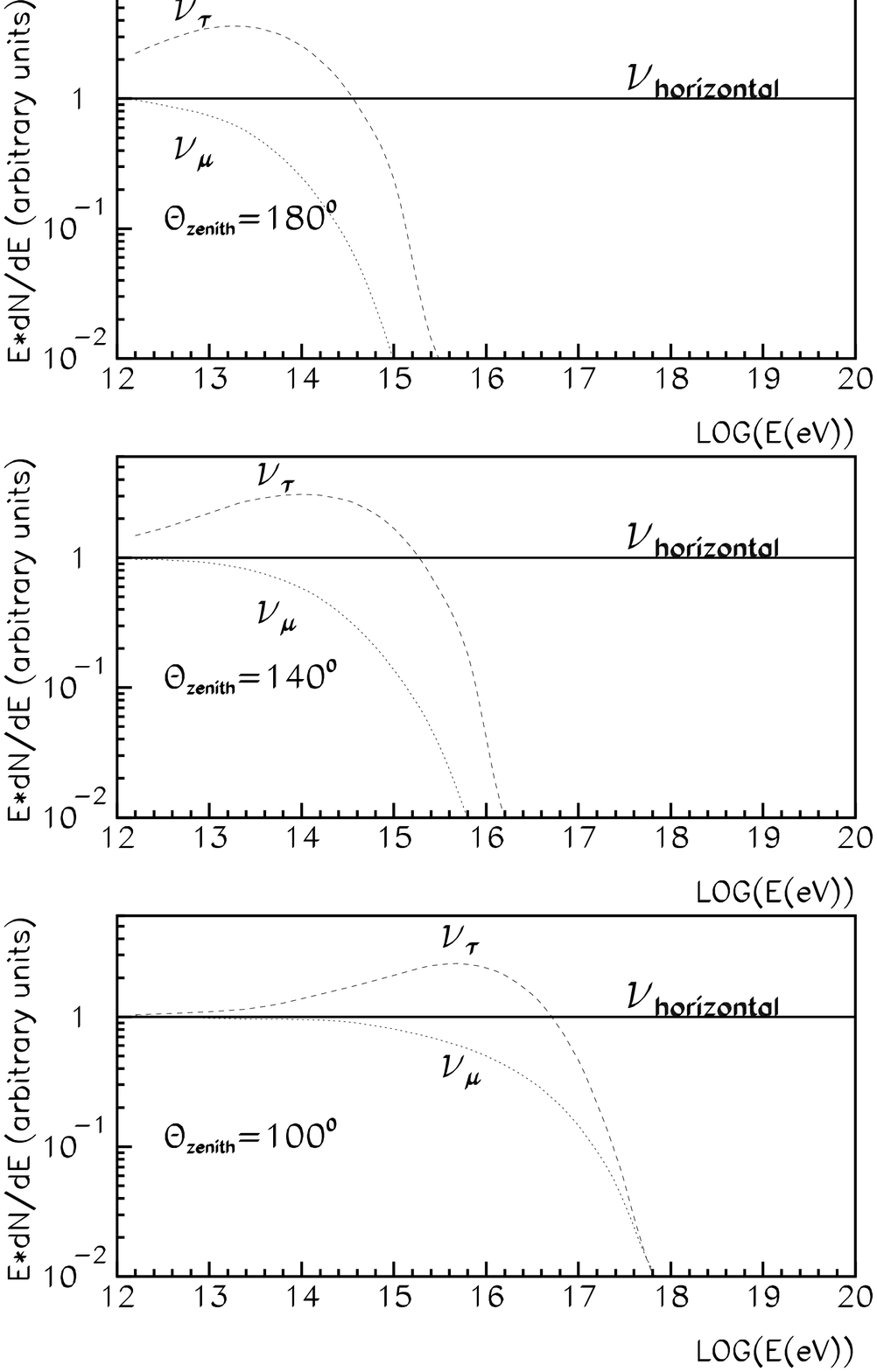,width=13cm}
\caption{Final energy spectra of \nut's and \num's crossing the Earth at
zeinth angles of 180$^o$, 140$^o$ and 100$^o$ for an initial energy spectrum 
$\propto E^{-1}$.}
\label{fig:fin1}
\end{center}
\end{figure}

\begin{figure}
\begin{center}
\epsfig{figure=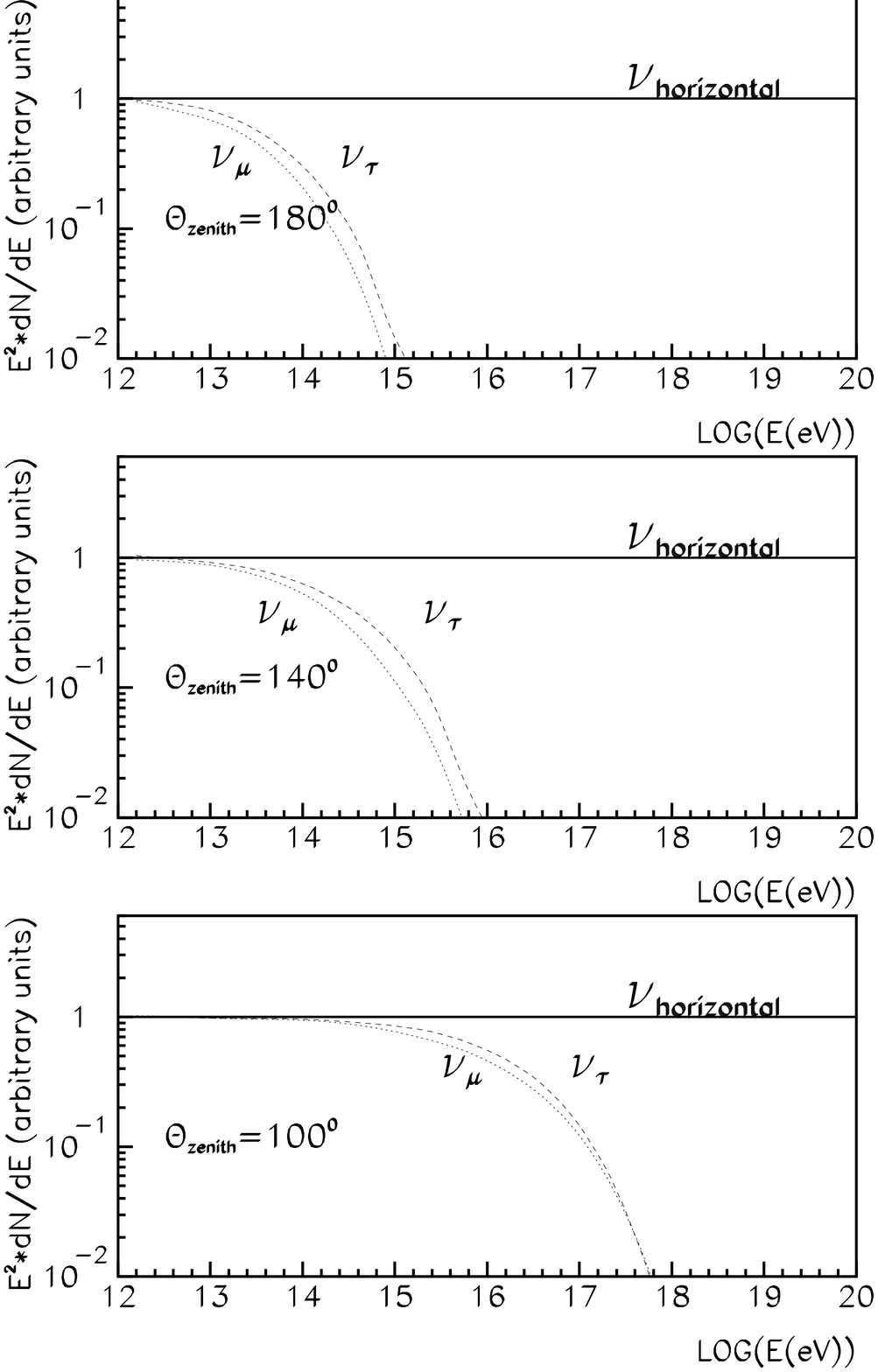,width=13cm}
\caption{Final energy spectra of \nut's and \num's crossing the Earth at
zeinth angles of 180$^o$, 140$^o$ and 100$^o$ for an initial energy spectrum 
$\propto E^{-2}$.}
\label{fig:fin2}
\end{center}
\end{figure}

\begin{figure}
\begin{center}
\epsfig{figure=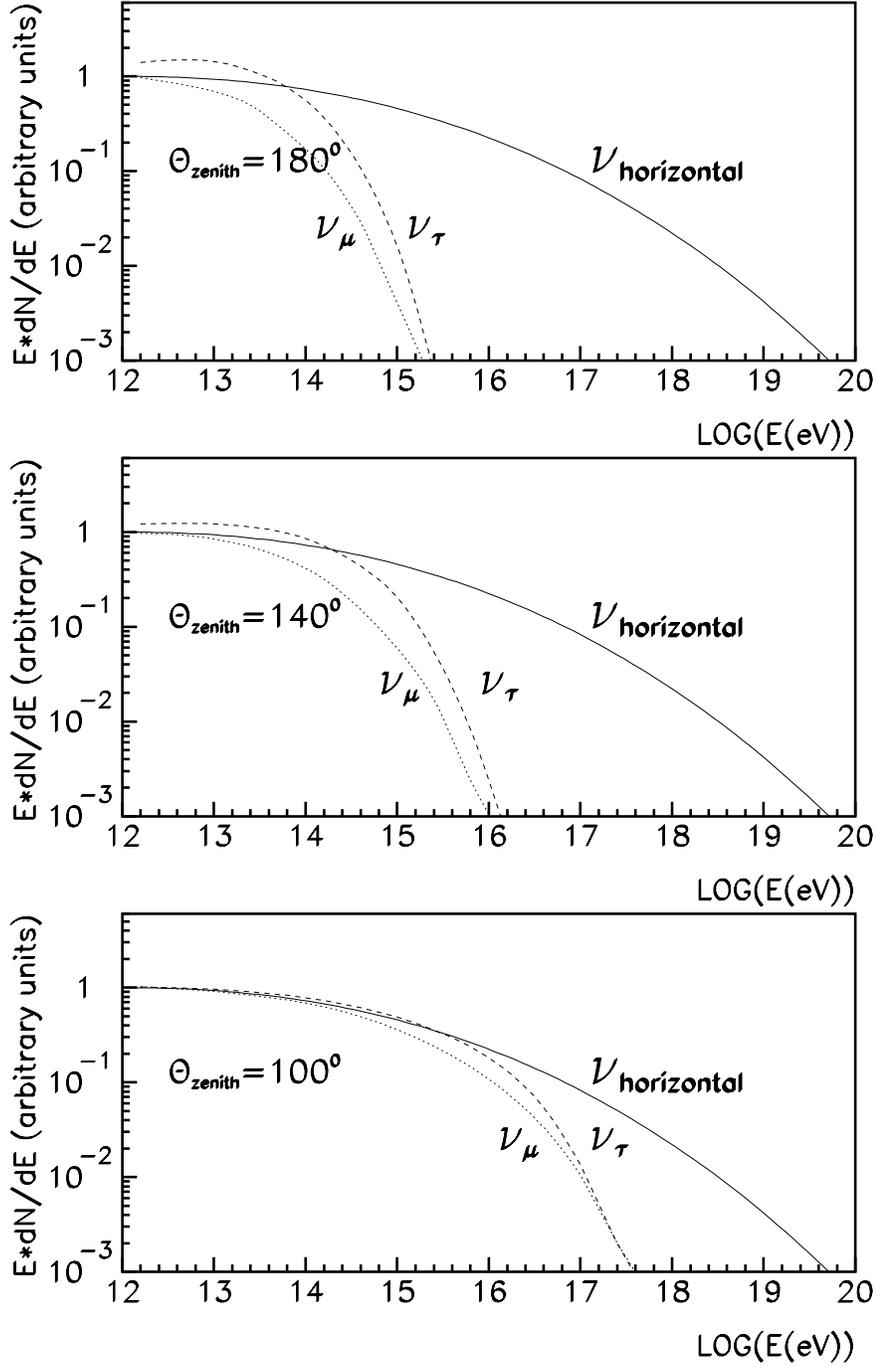,width=13cm}
\caption{Final energy spectra of \nut's and \num's crossing the Earth at
zenith angles of 180$^o$, 140$^o$ and 100$^o$ for an initial energy spectrum 
calculated for topological defect decays in ref.~\cite{defects}.}
\label{fig:fin3}
\end{center}
\end{figure}

In Figs.~\ref{fig:scpl1}, \ref{fig:scpl2} scatter plots of initial versus final energy are
shown. The initial energy is the energy of the neutrino before crossing the
Earth while the final energy is the energy of neutrino emerging from the Earth. 
The initial energy distribution has been chosen to be proportional to
$E^{-1}$ so to have a uniform distribution in $\log E$. Only muon neutrinos 
either interacting with neutral current or
not interacting at all are able to cross the Earth whereas the rest is absorbed
in that the muon arising from a charged current interaction has a radiation 
length much lower than its decay length at any considered energy.\\
On the other hand, the $\tau$ generated from a \nut charged current interaction
loses energy until its decay length becomes comparable to its radiation 
length, thereafter it decays into a new \nut. Unlike for muons, this occurs around 
$10^{16}-10^{17}\,$ eV and, as a consequence, CC interacting \nut's with 
$E_{\tau}\geq10^{17}\,$ eV, instead of being absorbed, populate the 
$E_{\tau} \leq 10^{17}$ eV energy region. The result of our Monte Carlo calculation 
shown in Fig.~\ref{fig:scpl2} is qualitatively in a fairly good agreement with a 
similar plot in ref.~\cite{hal98}, though a more quantitative comparison could 
be desirable.
The \nut propagation develops in several $\nu_{\tau}-\tau$ transformation steps.
The accumulation of \nut energies below $10^{17}\,$ eV is modulated by the zenith 
angle of the emerging particle because particle traverse different amounts of matter.
In Fig.~\ref{fig:scpl3} a scatter plot of final energy versus $\theta_{zenith}$
is shown. The input energy spectrum has been chosen once again $\propto E^{-1}$ while 
the zenith angle distribution is uniform in order to better display 
of the final energy angular dependence. The final energies cluster around 
$0.5\cdot 10^{17}\,$ eV for nearly horizontal particles and go down to $10^{14}\,$ eV
for vertically incoming particles.\\
We have studied the final energy spectrum of tau and muon neutrinos crossing the 
Earth at three different incident zenith angles for three different initial energy 
spectra: $\propto E^{-1}$ (see Fig.~\ref{fig:fin1}), $\propto E^{-2}$ 
(see Fig.~\ref{fig:fin2}) 
and a possible energy spectrum predicted for topological defect decays~\cite{defects}
(see Fig.~\ref{fig:fin3}).   
The main feature of \nut spectrum is indeed the presence of a characteristic 
bump before a steep drop, due to the \nut regeneration mechanism described
above. This clearly shows up for hard initial spectra (see Fig.~\ref{fig:fin1}) 
whereas it is hardly recognizable for softer initial spectra (see Fig.~\ref{fig:fin3})
and becomes a smooth slope in case of steeply falling energy spectra 
(see Fig.~\ref{fig:fin2}). The bump position strongly depends on zenith angle 
because energy loss increases as a function of encountered matter thickness.
The results obtained in this work are in good agreement with those 
by S. Iyer et al.~\cite{Iyer}, who took an analytical approach. 
The main difference between the two results is an excess of
\nut's around the characteristic bump in the $\propto E^{-1}$
energy spectrum in our work. This can be easily explained by taking into
account the difference in the initial neutrino spectrum. We have used 
a pure $\propto E^{-1}$ spectrum up to $10^{20}\,$ eV while S. Iyer 
et al. used a modulated cutoff for energies greater than $10^{17}\,$ eV. 
Since in a $\propto E^{-1}$ spectrum the number of neutrinos which are 
shifted from high energy to the characteristic bump at lower energy receives
approximately an equal contribution for each energy decade above the bump, the 
agreement is restored. Furthermore, in ref.~\cite{Iyer}, $\tau$ energy 
loss is not taken into account. However, the effect of such approximation
is almost negligible because in the energy region considered therein, 
below $10^{15}\,$ eV, $\tau$ decay lenght is much less than its interaction 
lenght.

\section{Conclusions}
 
Whilst, as it is well known, the Earth start to be opaque for muon neutrino 
propagation at energies around $10$ TeV, a regeneration mechanism relevant
to \nut prevents them from being absorbed. This mechanism is ultimately owing 
to the shorter, with respect to muon, $\tau$ lepton decay length. Nevertheless, 
the unavoidable radiative $\tau$ energy loss sets an upper bound on the energy 
of emerging \nut's to $10^{17}\,$ eV. Therefore, for underwater-underice 
neutrino telescopes, whose sensitivity will cover the whole neutrino spectrum,
\nut's could be the major source of neutrino events above the PeV region. 
In such detectors the charactestics of Fig.~\ref{fig:fin1} could be also used as
one of the signatures of tau neutrinos. For atmospheric fluorescence detectors 
the effective cut-off of $10^{17}\,$ eV could be below their energy threshold. 
If this is the case, those detectors should be able to detect only almost horizontal
neutrinos.

\section*{Acknowledgments}

We are very grateful to V. Naumov for bringing our attention to the problem
of \nut propagation and for very fruitful discussions.

\newpage

\newpage

\newpage

\newpage

\newpage

\newpage

\newpage

\end{document}